*Category*

# A powerful and efficient set test for genetic markers that handles confounders

Jennifer Listgarten[1,*], Christoph Lippert[1,*], Eun Yong Kang[1], Jing Xiang[1], Carl M. Kadie[1] and David Heckerman[1,*]

[1]eScience Group, Microsoft Research, Los Angeles, USA.

**ABSTRACT**

**Motivation:** Approaches for testing sets of variants, such as a set of rare or common variants within a gene or pathway, for association with complex traits are important. In particular, set tests allow for aggregation of weak signal within a set, can capture interplay among variants, and reduce the burden of multiple hypothesis testing. Until now, these approaches did not address confounding by family relatedness and population structure, a problem that is becoming more important as larger data sets are used to increase power.

**Results:** We introduce a new approach for set tests that handles confounders. Our model is based on the linear mixed model and uses two random effects—one to capture the set association signal and one to capture confounders. We also introduce a computational speedup for two-random-effects models that makes this approach feasible even for extremely large cohorts. Using this model with both the likelihood ratio test and score test, we find that the former yields more power while controlling type I error. Application of our approach to richly structured GAW14 data demonstrates that our method successfully corrects for population structure and family relatedness, while application of our method to a 15,000 individual Crohn's disease case-control cohort demonstrates that it additionally recovers genes not recoverable by univariate analysis.

**Availability:** A Python-based library implementing our approach is available at http://mscompbio.codeplex.com

**Contact:** {jennl,lippert,heckerma}@microsoft.com

## 1 INTRODUCTION

Traditional Genome-Wide Association Studies (GWAS) test one single nucleotide polymorphism (SNP) at a time for association with disease, overlooking interplay between SNPs within a gene or pathway, missing weak signal that aggregates in sets of related SNPs, and incurring a severe penalty for multiple testing. More recently, sets of SNPs have been tested jointly in a gene-set enrichment style approach (Holden *et al*, 2008), and also in seeking association between rare variants within a gene and disease (Wu *et al.*, 2011; Bansal *et al.*, 2010). As next generation sequencing rapidly becomes the norm, these set-based tests, complementary to single SNP tests, will become increasingly important. However, existing methods for testing sets of SNPs do not handle confounding such as arises when related individuals or those of diverse ethnic backgrounds are included in the study. Such confounders, when not accounted for, result in loss of power and spurious associations (Balding, 2006; A. Price *et al.*, 2010). Yet it is precisely these richly structured cohorts which yield the most power for discovery of the genetic underpinnings of complex traits. Moreover, such structure typically presents itself as data cohorts become larger and larger to enable the discovery of weak signals.

In this paper, we introduce a new, powerful and computationally efficient likelihood ratio-based set test that accounts for rich confounding structure. We demonstrate control of type I error as well as improved power over the more traditionally used score test. Finally, we demonstrate application of our approach to two real GWAS data sets. Both data sets showed evidence of spurious association due to confounders in an uncorrected analysis, while application of our set test corrected for confounders and uncovered signal not recovered by univariate analysis. Finally, our test is extremely computationally efficient owing to development of a new linear mixed model algorithm also presented herein, which makes possible, for example, set analysis of the 15,000 individual Wellcome Trust Case Control Consortium (WTCCC) data.

Several approaches have been used to jointly test sets of SNPs: *post hoc*, gene-set enrichment in which univariate *P* values are aggregated (Holden *et al*, 2008), operator-based aggregation such as "collapsing" of SNP values (Braun and Buetow, 2011; B. Li and Leal, 2008), multivariate regression, typically penalized (Schwender *et al.*, 2011; Malo *et al.*, 2008), and variance component (also called kernel) models such as a linear mixed models (Wu *et al.*, 2011, 2010; Quon *et al.*, 2013).

Our approach is based on the linear mixed model (LMM) which can equivalently be viewed as a multivariate regression. In particular, use of a LMM with a specific form of genetic similarity matrix is equivalent to regressing those SNPs used to estimate genetic similarity on the phenotype (Hayes *et al.*, 2009; Listgarten *et al.*, 2012). If one uses only SNPs to be tested in the similarity matrix as in Wu *et al.*, 2010, 2011, then one is effectively performing a multivariate regression test. However, by also using SNPs that tag confounders in a separate similarity matrix, our model can additionally correct for confounders, as has been done in a single-SNP test GWAS setting (Kang *et al.*, 2010; Yu *et al.*, 2006; Listgarten *et al.*, 2012; Lippert *et al.*, 2011). Finally, our approach allows one to condition on other causal SNPs, by way of the similarity matrix, for increased power, again, as has been done in single-SNP test setting (Atwell *et al.*, 2010; Listgarten *et al.*, 2012; Segura *et al.*, 2012).

---

[*]To whom correspondence should be addressed, equal contributions.

The use of LMMs to correct for confounders in genome-wide association studies (GWAS) is now widely accepted, because this approach has been shown capable of correcting for several forms of genetic relatedness such as population structure and family relatedness (Kang *et al.*, 2010; Astle and Balding, 2009; Yu *et al.*, 2006; A. Price *et al.*, 2010). Independently, the use of LMMs to jointly test rare variants has become prevalent (Wu *et al.*, 2010, 2011). In our new approach, we marry the aforementioned uses of LMMs to perform set tests in the presence of confounders within a single, robust, and well-defined statistical model.

Because of the aforementioned equivalence, our approach can also be viewed as a form of linear regression with two distinct sets of covariates. The first set of covariates consists of SNPs that correct for confounders (and other causal SNPs), that is, those which predict race and relatedness, for example. Inclusion of these SNP covariates makes the data for individuals independently and identically distributed (*i.e.*, knowing the value of these SNPs induces a common distribution from which the individuals are drawn). The second set of covariates consists of SNPs for a given set of interest, such as those SNPs belonging to a gene. We call our approach FaST-LMM-Set.

Computing the likelihood for our model—a LMM with two random effects—is, naively, extremely expensive, as it scales cubically with the number of individuals (*e.g.,* Listgarten *et al*, 2010). For example, on the 15,000 individual WTCCC data set we analyse, currently available algorithms would need to compute and store in memory genetic similarity matrices of dimension $15{,}000 \times 15{,}000$ and repeatedly perform cubic operations on them to test just a single set of SNPs—a practically infeasible approach. However, extending our previous work that made LMMs with a single random effect linear in the number of individuals (Lippert *et al* 2011) to the two-variance component model needed here, we bypass this computational bottleneck, yielding a new, two-random-effects algorithm which is linear in the number of individuals. This advance enables us to analyse data sets which could not otherwise be practically analysed, such as the 15,000 individual WTCCC cohort (The Wellcome Trust Case Control, 2007). As a case in point, using the naïve cubic approach to test the gene set IL23R (containing 14 SNPs) took 13 hours as compared to one minute for our new approach (all on a single processor), demonstrating a speedup factor of 780 (and significantly less memory usage because the genetic similarity matrix need never be computed with our approach).

## 2 METHODS

Let $N(v|u; \Sigma)$ denote a multivariate Normal distribution in $v$ with mean $u$ and covariance matrix $\Sigma$. The log likelihood of a one-variance-component linear mixed model in the linear regression view is given by

$$LL \equiv \log \int N\left(y|X\beta + \frac{1}{\sqrt{s}}Vw; \sigma_e^2 I\right) \cdot N(w|0; \sigma_g^2 I) \, dw,$$

where $y$ is a $1 \times N$ vector of phenotype values for $N$ individuals; $\beta$ is the set of the fixed effects of the covariates stored in the design matrix $X$; $I$ is an $N \times N$ identity matrix; $\sigma_e^2$ is the residual variance in the regression; $w$ are the $1 \times N$ random effects for the SNPs stored in the design matrix $V$ (dimension $N \times s$), and $N(w|0; \sigma_g^2 I)$ is the distribution for the weight parameters. That is, the random regression weights, $w$ are marginalized over independent Normal distributions with equal variance $\sigma_g^2$.

Equivalently, and more typically, the log likelihood is written with random effects marginalized out,

$$LL = \log N(y|X\beta; \sigma_e^2 I + \sigma_g^2 K),$$

where the genetic similarity (called the kernel in some contexts), $K$, is given by $K = \frac{1}{s}VV^T$, as is the case, for example, when $K$ is the realized relationship matrix (RRM) (B. J. Hayes *et al*, 2009; Lippert *et al*, 2011). Given this equivalence, the SNPs used to estimate genetic similarity (those in $V$) can be interpreted as a set of covariates in the regression.

In our model, we partition the random effects into two sets: one set of random effects, $w_C$ (with design matrix $V_C$), are used to correct for confounders (and condition on causal SNPs) using $s_c$ SNPs, while the other set, $w_S$, are used to test the $s_s$ SNPs of interest in the corresponding design matrix, $V_S$. The log likelihood (in the linear regression view) is then written

$$LL = \log \iint N\left(y|X\beta + \frac{V_C}{\sqrt{s_c}}w_C + \frac{V_S}{\sqrt{s_s}}w_S; \sigma_e^2 I\right) \cdot N(w_C|0; I) \, N(w_S|0; I) \, dw_C dw_S,$$

where each set of random effects has a separate variance ($\sigma_C^2$ and $\sigma_S^2$). Again, we can equivalently write this in the marginalized form,

$$LL = \log N\left(y|X\beta; \sigma_e^2 I + \frac{\sigma_C^2}{s_c}V_C V_C^T + \frac{\sigma_S^2}{s_s}V_S V_S^T\right).$$

For convenience, we re-parameterize this as

$$LL = \log N(y|X\beta; \sigma_e^2 I + \sigma_g^2[(1-\tau)K_C + \tau K_S]), \quad (1)$$

where now the covariance matrix, $K$, has been partitioned into two variance components:

$$K \equiv (1-\tau)K_C + \tau K_S,$$

using $K_C \equiv \frac{1}{s_c}V_C V_C^T$ ($V_C$ is of dimension $N \times s_c$) to account for confounders, and $K_S \equiv \frac{1}{s_s}V_S V_S^T$ ($V_S$ is of dimension $N \times s_s$) to model signal from a pre-defined set of SNPs of interest, such as those within a gene. The scalar parameter $\tau \in [0,1]$ is estimated from the data by, for example, restricted maximum likelihood (REML). The null model for our set test is given by $\tau = 0$, while the alternative model allows $1 \geq \tau \geq 0$.

Until lately, estimating the parameters and computing the likelihood of a LMM was cubic in the number of individuals. However, we have recently shown that when the number of SNPs used to estimate genetic similarity, $s$, is less than the cohort size, $N$, and when genetic similarity matrix, $K$, factors as $VV^T$ ($V$ of dimension $N \times s$), then the computations (and memory requirement) become linear in $N$ (Lippert *et al*, 2011). So far this result has been applied in the context of correcting for confounders in a univariate GWAS with just a single variance component. Originally the $s$ SNPs for inclusion in $\boldsymbol{V}$ were obtained by sampling SNPs genome-wide and relying on linkage disequilibrium (Lippert *et al*, 2011). However, in light of the equivalence with linear regression, it became clear that one should choose the SNPs with feature selection as one would do in any statistical modelling problem (Listgarten *et al.*, 2012, 2013; Lippert, Quon, *et al.*, 2013). For example, one can do an uncorrected, univariate scan of the SNPs to select those which should be used to correct for confounders (and causal SNPs to condition on), and this is precisely the approach we take here, just as in (Listgarten *et al.*, 2012; Lippert, Quon, *et al.*, 2013).

Thus, in our approach, we select $s_c$ SNPs for $\boldsymbol{K_C}$ by first sorting all available SNPs according to their univariate linear-regression *P* values (in increasing order), and then evaluate the use of more and more SNPs in order until we find an optimal number of SNPs (Listgarten *et al.*, 2012). This resulted in 650 and 310 SNPs for the GAW14 and Crohn's analyses, respectively. Additionally, any SNPs that were being tested (*i.e.,* those in $\boldsymbol{V_S}$), and those within 2 centimorgans, were removed from $\boldsymbol{V_C}$ so as not to contaminate the null model (Listgarten, Lippert, C. M. Kadie *et al*, 2012). This type of approach for correction of confounders in univariate tests has previously been demonstrated to work well on a broad range of data sets (Listgarten, Lippert, C. M. Kadie *et al*, 2012; Listgarten, Lippert, and Heckerman, 2012). However, we note that on some other data sets, we found that our original approach (Listgarten *et al.*, 2012) could be sub-optimal; therefore, we have now moved to a slightly different approach (Lippert, Quon, *et al.*, 2013).

For estimation of variance parameters and computation of the likelihood ratio test statistic, we use REML, which is itself a valid likelihood for the likelihood ratio test (LRT) and can be computed in the same time and memory complexity as the (unrestricted) likelihood. Details on efficient parameter estimation are provided in (Lippert *et al.*, 2011).

When testing sets in an uncorrected manner, that is, without accounting for confounders (which we did for comparison purposes), we omitted the portion of the variance which corrected for confounders, $\boldsymbol{K_C}$. In particular, we set $\tau = 1$, and tested the significance of $\sigma_g^2$ with the same LRT described next.

*P* Value Computation

We have now fully specified our model for doing set tests when confounders are present. To obtain a *P* value on the set of SNPs of interest, such as those belonging to a gene (*i.e.,* those in $\boldsymbol{V_S}$), we use an LRT. In particular, to test the significance of the set of SNPs of interest, we compare the maximum restricted likelihood of the data with and without the set of SNPs of interest, that is, the maximum restricted likelihood of the alternative and null models. More formally, our null hypothesis is given by $H_0$: $\tau$=0, while our alternative hypothesis is given by $H_a$: $1 \geq \tau \geq 0$.

To obtain calibrated *P* values, we require an accurate estimate of the distribution of statistics under the null hypothesis. However, obtaining a sufficiently accurate estimate of this distribution is not straight-forward. Standard software uses a parametric form for this distribution of $0.5\chi_o^2 : 0.5\chi_1^2$—a 50-50 mixture of two $\chi^2$ distributions, the first with zero degrees of freedom, and the other with one degree of freedom. The former accounts for the fact that the tested parameter is on the boundary of the allowed space in the null model (Self and Liang, 1987; Dominicus *et al.*, 2006)—that is, to account for the fact that $\tau = 0$ in the null model, and $1 \geq \tau \geq 0$ in the alternative. The necessary regularity conditions for this null distribution to hold, include that the outcome variable can be partitioned into a large number of identically and independently distributed sub-vectors (Greven *et al.*, 2008)—conditions which are not generally met in our setting because individuals may be arbitrarily related to one another. It has been shown that when the regularity conditions are not met, the $0.5\chi_o^2 : 0.5\chi_1^2$ distribution yields conservative *P* values (Greven *et al.*, 2008) because the mixing weight on the $\chi_o^2$ component is too low at fifty percent. We have also found this to be the case in our setting (Table 1).

Although one might consider use of a parametric bootstrap to estimate the null distribution, *e.g.,* (Greven *et al.*, 2008), such an approach dramatically increases the running time over computation of the test statistics themselves because of the extremely large number of bootstrap test statistics needed. Yet another alternative is to use an empirical distribution based on permutations, which faces a similar problem. However, one can use many fewer permutations by instead assuming a parametric form of the null distribution and then fitting the few required parameters to the test statistics generated from the permutations (Lee, Emond, *et al.*, 2012). It is such an approach that we take here. Note that this approach assumes that the null distribution of test statistics is the same across all tests, an assumption that has also been made in the small sample correction in SKAT-O and elsewhere (Lee, Emond, *et al.*, 2012; Greven *et al.*, 2008; Greven, 2007).

The parametric form of the null distribution that we assume, and to which we fit null distribution test statistics to, is inspired by (Greven, 2007; Greven *et al.*, 2008), who reported that a mixture of $\chi_o^2$ and $a\chi_1^2$, where $a$ is the scaling parameter for the scaled chi-square distribution, yielded good type I control when testing a variance component in a single-component LMM. We use the same parametric form of the null distribution, except, to gain additional flexibility for the two-component LMM, we allow the degrees of freedom on the second component, $d$, to be different from 1 (finding this to be useful in the sense that we estimate $d \neq 1$). That is, we use the null distribution, $\pi\chi_o^2$:$(1 - \pi)a\chi_d^2$ with free parameters $\pi$, $a$, and $d$. Using this distributional form, we found that a fit of the free parameters to the (full collection of test) statistics yielded *P* values that were too liberal in the tail (Table 1). Thus, we instead fit our parametric parameters using only the most significant tail of the null distribution of test statistics—in particular the top 10% of null test statistics. (Note that in our experiments with just a single variance component, *P* values were also liberal in the tail—those for which p<<0.05. This regime was not examined by Greven *et al.*).

We now describe the details of our approach for estimating the free parameters, $\pi$, $a$, and $d$, of this null distribution. To generate a single null test statistic for a set, we permuted the individuals for only the SNPs in that set. Because we do not permute the SNPs (rather, the individuals), the pattern of linkage disequilibrium between the SNPs within a single test remains intact. Although we permute the individuals, who are not (generally) identically and independently distributed, we do so only for the SNPs in the test set, leaving any confounding signal among the covariates, the confounding SNPs and the phenotype intact. We found that null distribution parameter estimates stabilized with the use of 10 permutations per test (for both WTCCC and GAW14). Thus, our procedure has a runtime roughly a factor of ten larger than if we had not needed permutations. Within a gene, we use the same permutation for all SNPs, and we used the same 10 permutations across all sets.

Given this permutation-generated sample of test statistics from the null distribution, we fit the parameters $\pi$, $a$, and $d$ as follows. The $\chi_o^2$ distribution is a Dirac delta function at 0—that is, this component of the null distribution yields only test statistics of 0, and correspondingly p=1. Furthermore, the $\chi_{d>0}^2$ yields a test statistic of 0 with measure zero. Consequently, one can obtain good estimates of the parameters simply by assuming that precisely those tests with variance parameter estimate $\tau = 0$ belong to the $\chi_o^2$ component, and then estimating $\pi$ as the proportion of tests belonging to this component. We then estimate $a$ and $d$ directly from the non-zero test statistics (those likely to belong to the $a\chi_d^2$ component) using a regression in which these parameters are adjusted such that resulting LRT *P* values have the least squared error with the theoretical *P* values (derived conditionally on an estimate of $\pi$). Specifically, we use the log *P* value squared error, and only use the smallest 10% of *P* values in the regression. This truncated regression approach consistently yielded calibrated quantile-quantile plots (Figure 1) and also controlled type I error (Table 1). Furthermore, it yielded better power than the score test (Table 2).

**3**

In summary, our overall approach is as follows: (1) for each set to be tested, permute the individuals of the SNPs belonging to this set, all in the same manner; (2) compute the restricted LRT statistic for this permuted data to obtain a test statistic from the null distribution; (3) repeat step 1 ten times; (4) estimate the proportion of test statistics drawn from the $\chi_0^2$ component, $\hat{\pi}$, as the proportion of tests in which the parameter $\tau=0$; (5) use the largest 10% of test statistics to perform a regression to fit the $a\chi_1^2$ component —that is find $\hat{a}$ and $\hat{d}$ which minimize the squared error of the $\log_{10} P$ values with their theoretical values (uniform distribution on $[\hat{\pi}, 1]$); (6) compute the test statistic for all sets (non-permuted data) and then compute the corresponding $P$ values for these using the null distribution $\hat{\pi}\chi_0^2 : (1 - \hat{\pi})\hat{a}\chi_{\hat{d}}^2$.

In application to real data (described next), our procedure yielded $\pi = 0.641$, $a = 2.29$, $d = 0.961$, on the GAW14 data, and $\pi = 0.643$, $a = 1.41$, $d = 0.85$ on the WTCCC data.

Data Sets and Other Methods

The first data set was obtained from the Genetic Analysis Workshop (GAW) 14 (Edenberg *et al.*, 2005). It consisted of autosomal SNP data from an Affymetrix SNP panel and a phenotype indicating whether an individual smoked a pack of cigarettes a day or more for six months or more. The cohort included over eight ethnicities and numerous close family members—1,034 individuals in the dataset had parents, children, or siblings also in the dataset. In addition to the curation provided by GAW, we excluded a SNP when either (1) its minor allele frequency was less than 0.05, (2) its values were missing in more than 10% of the population, or (3) its allele frequencies were not in Hardy-Weinberg equilibrium ($p < 0.001$). In addition, we excluded an individual when more than 10% of SNP values were missing. After filtering, there were 7,579 SNPs across 1,261 individuals.

The second data set comprised the Wellcome Trust Case Control Consortium (WTCCC) 1 data and consisted of SNP and phenotype data for seven common diseases: bipolar disorder, coronary artery disease, hypertension, Crohn's disease, rheumatoid arthritis (RA), type-I diabetes, and type-II diabetes (The Wellcome Trust Case Control, 2007). Each phenotype set contained about 1,900 individuals. In addition, the data included a set of approximately 1,500 controls from the UK Blood Service Control Group (NBS). The data did not include a second control group from the 1958 British Birth Cohort (58C), as restrictions on it precluded use by a commercial organization. Our analysis for the Crohn's phenotype used data from the NBS group and the remaining six phenotypes as controls (Lippert, Listgarten, *et al.*, 2013). We filtered SNPs as described by the WTCCC (Laaksovirta *et al.*, 2010), and additionally excluded a SNP if either its minor-allele frequency was less than 1%, it was missing in greater than 1% of individuals, or its genetic distance was unknown. After filtering, 356,441 SNPs remained. Unlike the approach used by the WTCCC, we included non-white individuals and close family members to increase the potential for confounding and thereby better exercise the LMM. In total, there were 14,925 individuals across the seven phenotypes and control, as in our previous work (Listgarten *et al.*, 2012; Lippert *et al.*, 2011; Lippert, Listgarten, *et al.*, 2013). We concentrated our evaluations on Crohn's disease, as inflation for this phenotype was greatest with an uncorrected univariate analysis.

For the WTCCC data, we grouped SNPs into gene sets using gene positions provided on the USCSC Genome Browser (http://genome.ucsc.edu/) (Kent *et al.*, 2002; Dreszer *et al.*, 2012) using build hg19 (we also converted the original WTCCC annotations to this build), which yielded 13,850 gene sets. Because the GAW14 SNPs mapped to only 251 non-singleton gene sets with this strategy, we instead formed sets for this data set by using overlapping 1 centimorgan windows, yielding 2,157 sets. For WTCCC, the set sizes ranged from 1 to 748, with a mean value of 11, and a standard deviation of 24. For the GAW14 data, the set sizes ranged from 2 to 38, with a mean value of 5, and a standard deviation of 4. More generally, this approach of forming sets from windows of nearby SNPs along the genome could be used to map an entire genome into sets, even when the SNPs do not lie in genes. However, it is not our goal here to evaluate different ways in which one might group SNPs, but instead to demonstrate that we can test sets of SNPs in the presence of confounders.

All analyses assumed additive effects of a SNP on phenotype, using a 0/1/2 encoding for each SNP (indicating the number of minor alleles for an individual). Missing SNP data was mean imputed. Multiple testing was accounted for with a Bonferroni correction.

In counting hits for Crohn's disease (Table 4), we omitted any genes found in the MHC region because this region is complicated by very long-range linkage disequilibrium. We used positions 29-34 Mb on chromosome 6 as the boundaries of the MHC, as suggested by the MHC sequencing consortium (Pereyra *et al.*, 2010).

Experimental Setup to Assess Control of Type I Error and Power

We used synthetic data based on the real WTCCC data to assess the quality of our new method, as well as to compare it against a score test. In particular, to assess type I error, we used all SNPs from the WTCCC data set, and then permuted the individuals for SNPs in each set tested so as to create null-only test statistics. We permuted the data in this way a total of 72 times, yielding 997,200 null test statistics (because 13,850 sets were tested for each data set). We additionally permuted another 10 data sets in order to estimate the parameters of the null distribution $(\pi, a, d)$ as prescribed by our approach.

For assessment of power, we again used all SNPs from the WTCCC data set, and then generated synthetic phenotypes using a linear mixed model. To do so, we first we fit the null model to the real data to obtain estimates of the parameters $\sigma_e^2$ and $\sigma_g^2$. Then we used the model $p(y) = N(y|0; \sigma_c^2 K_C + \sigma_s^2 K_S])$, with confounding variance $\sigma_c^2$ equal to estimated environmental noise, $\sigma_e^2 = 0.094$, and genetic variance, $\sigma_s^2$, equal to the estimated genetic variance $\sigma_g^2 = 0.0125$. Furthermore, we used the same 310 confounding SNPs for $K_C$ as used on the real data, while using all 321,839 SNPs further than 2 centimorgans away from those in $K_C$ as the causal SNPs for $K_S$ (those contained in the true positive sets in our power experiments). We generated five phenotypes in this way. The resulting phenotypes behaved much like the real data in that, on average, we found 10 Bonferroni-corrected sets on each of five data sets, as compared to the 23 found on the real data (note that Table 4 does not include SNPs from the MHC region and therefore shows only 16). For both type I error and power experiments, we tested the same gene sets as on the real data, except for power we did not include any gene sets containing SNPs in the $V_C$ (those SNPs used to correct for confounding) or within 2 centimorgans of these SNPs, to be sure that the sets were unambiguously true positives.

When comparing our LRT approach against a score-based test, we used the same score test as SKAT (Wu *et al.*, 2011) which uses the Davies method to compute $P$ values from the null distribution (but here with our FaST-LMM-Set model). This score test has previously been shown to control type I error (Lee, Emond, *et al.*, 2012), consistent with our own findings (not shown).

Linear-Time Computations

What remains left to explain is how to achieve the linear-time speedup in the present setting—the case of two random effects. The crux of the cubic to linear-time speedup in the single random effect model was to bypass construction of $K$ and the required spectral decomposition of $K$ by recognizing that one can instead use $V$ and the spectral decomposition of $V$ (Lippert *et al.*, 2011). Note that we can view the two-random-effects model as a single random effect with covariance $K \equiv (1 - \tau)K_C + \tau K_S$. To use the algebraic speed-up just mentioned, we observed that $K = VV^T$, where now

$$V \equiv \left[\frac{1}{\sqrt{s_c}} V_C (1-\tau)^{\frac{1}{2}} ; \frac{1}{\sqrt{s_s}} V_S \tau^{\frac{1}{2}}\right],$$

using $[U; W]$ to denote the side-by-side concatenation of matrices $U$ and $W$. So long as $s = s_s + s_c < N$, which was true for all of the data sets examined here (and almost all others we have analyzed), we obtained the linear-time computations and memory footprint just as in Lippert *et al.* (2011). (One might also consider using low rank update equations of the type used in Listgarten *et al.* (2012) to perform exclusion, although we have not yet implemented this.) Finally, to perform parameter estimation in this two-random-effects model, we used an approach similar to that reported in Lippert *et al.* (2011). That is, we used a one-dimensional Brent search optimization routine to find the value of $\tau$ which maximized the restricted likelihood. For each call to the restricted likelihood for a particular value of $\tau$, efficient computations were performed as in (Lippert *et al.*, 2011), except using the two random effects.

## 3 RESULTS

Type I Error and Power on Synthetic Data

First we examined whether our LRT approach controlled type I error. As described in Methods, we generated null-only test statistics by way of permutations on the WTCCC data, obtaining a total of roughly 1 million test statistics. The type I error was controlled (Table 1). Note that neither (1) fitting the null distribution parameters (to all test statistics) by way of maximum likelihood, nor (2) use of a $0.5\chi_o^2 : 0.5\chi_1^2$ null distribution, yielded calibrated $P$ values. The first was liberal, while the latter was conservative (Table 1). Finally, quantile-quantile plots in Figure 1 additionally demonstrates good calibration over the entire range of $P$ values from our method, for the same points as in Table 1. Because the score test we used has already been shown to control type I error (Lee, Emond, *et al.*, 2012), we do not report on it here, but note that we did find the same in our own experiments (not shown).

**Table 1**. Type I error estimates for FaST-LMM-Set using one million tests across various levels of significance, α.

| Significance Level | $\alpha = 10^{-5}$ | $\alpha = 10^{-4}$ | $\alpha = 10^{-3}$ |
|---|---|---|---|
| Fast-LMM-Set | $1 \times 10^{-5}$ | $1.21 \times 10^{-4}$ | $1.01 \times 10^{-3}$ |
| non-truncated ML | $2 \times 10^{-5}$ * | $1.83 \times 10^{-4}$ * | $1.26 \times 10^{-3}$ * |
| $0.5\chi_o^2 : 0.5\chi_1^2$ | $5 \times 10^{-6}$ | $4 \times 10^{-5}$ * | $4.55 \times 10^{-4}$ * |

The first row shows results for our LRT-based method; the second row ("non-truncated ML") shows results when fitting the null distribution parameters using maximum likelihood with all test statistics; the third row shows results using a $0.5\chi_o^2 : 0.5\chi_1^2$ null distribution. Results significantly different from expected according to the binomial test (p<0.05) are denoted with a *.

Next we compared the power of our LRT approach to a score test approach (both using the same two random effects model) on synthetic data (see Methods). Over five synthetic data sets and a range of significance levels, LRT found significantly more sets than the score test (Table 2). Furthermore, on the real WTCCC data, LRT again found significantly more sets (Table 4).

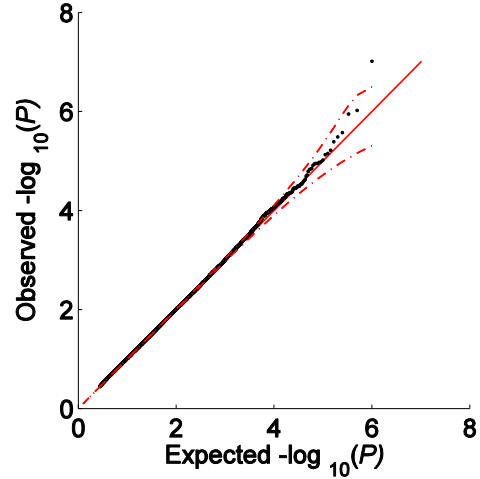

**Fig. 1.** Quantile-quantile plot of observed and expected $\log_{10} P$ values on the null-only WTCCC data sets (same data as used for Table 1) for FaST-LMM-Set. Dashed red error bars denote the 99% confidence interval around the solid red diagonal. Points shown are for null-only data (generated by permuting individuals in the SNPs to be tested—see Methods) and only for the non-unity $P$ values (those assumed to belong to the non-zero degree of freedom component of the null distribution). The portion of the expected distribution of $P$ values shown is uniform on the interval $[\hat{\pi},1]$, where $\hat{\pi}$ is the estimated mixing weight in the null distribution.

**Table 2**. Power experiments.

| α | LRT | score | *P* value |
|---|---|---|---|
| $3.6 \times 10^{-6}$ | 44 | 26 | 0.03 |
| $10^{-5}$ | 60 | 39 | 0.03 |
| $10^{-4}$ | 172 | 138 | 0.05 |
| $10^{-3}$ | 556 | 509 | 0.14 |
| $10^{-2}$ | 2419 | 2195 | 0.0009 |

Number of tests with *P* values less than α. The last column shows the results of a binomial test comparing the number of tests found by LRT as compared to the score test. The first row denotes the Bonferroni threshold for the WTCCC data set.

Application to Real Data

We investigated our new approach on two data sets. The first was the Genetic Analysis Workshop (GAW) 14, which included data from over eight ethnicities and numerous close family members for a total of 1,261 individuals. After filtering there were 7,579 SNPs available for analysis. The second data set was from the WTCCC from which we used 14,925 individuals and 356,441 SNPs in our analysis. We used the Crohn's phenotypes because this was the one showing the most confounding in an uncorrected analysis. Unlike the WTCCC (The Wellcome Trust Case Control, 2007), we included non-white data for individuals and close family members to increase power and because the LMM can treat them properly (Kang *et al.*, 2010; Astle and Balding, 2009; A. Price *et al.*, 2010).

To judge the degree of confounding due to genetic relatedness, and to ensure that our LMM approach could sufficiently correct for confounding, we ran both an uncorrected and corrected *univariate* analysis on each data set, because this is a well-understood test that has been reported on before. Here, the extent of test statistic inflation due to unmodelled confounders was assessed using the λ statistic, also known as the inflation factor from genomic control (Devlin and Roeder, 1999). The value λ is defined as the ratio of the median observed to median theoretical test statistic. Values of λ substantially greater than (less than) 1.0 are indicative of inflation (deflation). As can be seen in Table 3, without correction, the test statistics appear to be inflated. Although one might consider $\lambda_{GC} = 1.08$ (seen on the corrected analysis of WTCCC) as still moderately inflated, it has been shown that complex, highly polygenic traits lead to increases in $\lambda_{GC}$ (Yang *et al.*, 2011) in the absence of spurious signal. Moreover, the WTCCC themselves reported $\lambda_{GC}$ in the range of 1.08-1.11 upon removal of individuals from different races and related individuals (neither of which we removed), and also upon adjustment with two principal components, suggesting that a $\lambda_{GC}$ of 1.08 is the result of polygenic influence (The Wellcome Trust Case Control, 2007).

**Table 3**. $\lambda_{GC}$ of univariate tests for confounding-corrected and naïve methods

| Method | GAW14 | WTCCC |
|---|---|---|
| Uncorrected | 3.80 | 1.30 |
| FaST-LMM | 1.01 | 1.08 |

FaST-LMM denotes a one-component (to correct for confounding) linear mixed model, testing one SNP fixed effect (Listgarten *et al.*, 2012); Uncorrected refers to no correction for confounding (linear regression).

Having established that both of our data sets required correction for confounders and that the LMM with our chosen background genetic similarity matrix, $K_S$, sufficiently corrected for confounders, we next applied FaST-LMM-Set, using the same LMM-correcting component as in the univariate test. The full set of results is available for all analyses in Supplemental Table 1. On GAW14, the uncorrected set analysis yielded 241 significant sets, whereas FaST-LMM-Set, which corrects for confounding, yielded none. It is thought that this data set contains little, if any signal (for example based on the univariate analysis). On WTCCC Crohn's disease, an uncorrected set analysis yielded 26 significant sets, whereas FaST-LMM-Set yielded 16 (Table 4). Next we investigate these sets in detail.

To validate the significant sets recovered on the WTCCC Crohn's phenotype we used a meta-analysis (Franke *et al.*, 2010; Listgarten *et al.*, 2012). If a set we found as significant was within 50 kilobases of a validated SNP/region, we counted it as a true positive). Additionally, for the genes not validated by the meta-analysis, we conducted a literature search. Detailed validation results are provided in Supplemental Table 1. Using our newly developed method, FaST-LMM-Set, we found 16 significant gene sets, of which all but one were validated by the meta-analysis. The remaining gene, SLC24A4, performs a similar function to the validated gene SLC22A4—both are cation transporters (www.genecards.org (Rebhan *et al.*, 1997)) —suggesting a promising candidate for follow-up.

**Table 4**. Validation of methods on WTCCC Crohn's disease

| Method | in meta-analysis | supported by literature | no support found |
|---|---|---|---|
| FaST-LMM-Set | 15 | 1 | 0 |
| FaST-LMM-Set-Score | 7 | 0 | 0 |
| FaST-LMM-Set (uncorrected) | 17 | 3 | 6 |

FaST-LMM-Set denotes our newly developed method which corrects for confounding and uses our LRT approach; FaST-LMM-Set (uncorrected) is the same but does not correct for confounding with a second variance component; FaST-LMM-Set-Score is the same as FaST-LMM-Set but uses a score test (as described in Methods) instead of an LRT. Columns: "in meta-analysis" shows the number of significant sets validated by a meta-analysis (Franke *et al.*, 2010); "supported by literature" denotes the number of significant sets found by a literature search; "no support found" denotes the number of sets supported neither by the meta-analysis nor a literature search.

Advantages of Set Tests over Univariate Tests

In the course of our analyses we noticed that some sets with small *P* values had almost no univariate signal in any of the SNPs. In particular, among the 16 sets in the WTCCC data supported by either meta-analysis or literature search, six (C1orf141, SAG, SLC24A4, SLC22A4, TCTA, and PTPN2) were missed by the univariate analysis (*i.e.*, a SNP lying within 50 kilobases of any of the regions reported by Frank *et al.* was not found to be significant). One of the motivations for doing set analysis is to uncover signals for such regions. The intuition here is the same as in a univariate conditional GWAS analysis. That is, conditioning on variables can lead to an increase in power, revealing signal that would be hidden without the conditioning (Atwell *et al.*, 2010; Segura *et al.*, 2012). Thus the set test acts not only to aggregate weak signal, but also to unmask signal hidden by covariates included by virtue of doing a set test. We investigated one such case in detail. In particular, we computed the univariate *P* values for each of the 15 SNPs associated with the gene SLC22A4, marginally, as well as conditioned on all the other SNPs in this gene, using a LMM to correct for confounding. This gene was found to be associated with Crohn's disease using FaST-LMM-Set with $p = 7.6 \times 10^{-8}$. The smallest marginal univariate *P* value was $1.2 \times 10^{-5}$, but when we conditioned on the other SNPs in the set, the smallest conditional univariate *P* value obtained was $7.0 \times 10^{-8}$. This result demonstrates the increased power afforded by the set test owing to the interplay of SNPs within the gene that is missed by a univariate approach.

Significance of Sets is Independent of Set Size

On data with phenotypic association, we expected that there could be correlation between set size and *P* value, because with a larger set, there could be more predictive SNPs and more power. Furthermore, we expected that when confounders were not properly accounted for in the set analysis, that the more SNPs in a set, the more power the set would have to detect these confounders, and therefore the stronger the correlation between set size and *P* value would appear. The correlations on our real data were consistent with these expectations. In particular, we saw no significant correlation for FaST-LMM-Set (which corrects for confounders) but significant correlation when we did not correct for confounders (Table 5).

We also expected that on null-only data, when confounders were properly accounted for, the set *P* value and set size would be independent. Consistent with this expectation, when we permuted the Crohn's phenotype to remove signal, the FaST-LMM-Set correlation was reduced to $\rho = 0.019$ ($p = 0.18$).

**Table 5**. Pearson correlation of $\log_{10}(P)$ values with set size.

| Method | FaST-LMM-Set (uncorrected) | FaST-LMM-Set |
|---|---|---|
| GAW14 | **0.27 ($2 \times 10^{-34}$)** | 0.001 (0.98) |
| WTCCC | **0.051 ($3 \times 10^{-5}$)** | 0.025 (0.06) |

FaST-LMM-Set denotes our newly developed method; FaST-LMM-Set (uncorrected) is the same but does not correct for confounding with a second variance component. The *P* value is reported in parentheses next to the value for $\rho$. Significant entries are bolded. We excluded P values from the zero degree-of-freedom component of our one-sided test, as their inclusion would violate the assumptions of the Pearson correlation test.

## 4 DISCUSSION

We have developed a novel, efficient approach for testing sets of genetic markers in the presence of confounding structure such as arises from ethnic diversity and family relatedness within a cohort. Application of this algorithm demonstrated that our method corrects for confounders and uncovers signal not recoverable by univariate analysis.

Although we did not analyse rare variant data, we have shown elsewhere that the underlying LMM methodology works well to correct for confounding due to rare variants in a univariate setting (Listgarten *et al.*, 2013). Furthermore, others have already shown that LMM-based set tests work well for detection of sets of associated rare variants (Wu *et al.*, 2011). It follows that the hybrid approach that we presented here is likely to prove effective in the setting of testing sets of rare variants in the presence of confounders, although this remains to be investigated fully. For example, we have found the use of a linear model on a case-control phenotype to yield inflated tests statistics when testing rare variants.

We have demonstrated that our LRT outperforms a score test for our model and setting. This is perhaps unsurprising given that the score test can be viewed as an approximation to the LRT by a second-order Taylor series expansion (Buse, 2007) in the neighbourhood of the null model. Furthermore, given its robust properties, the LRT is considered the benchmark for statistical testing (Crainiceanu, 2008). We note, however, that in some recent work (D.-Y. Lin and Z.-Z. Tang, 2011), when testing for rare variants using a logistic fixed effects model, a score test was found to perform better than LRT, which was found to be liberal. Although the best test may depend on context, we note that Lin *et al.* used a different model than we did and, in particular, did not use a variance component approach. Also, they used closed-form, asymptotic-based LRT *P* values rather than making use of empirically-derived null distributions as we have done here.

For many cases of hidden structure in genetic data, the use of principal component-based covariates is sufficient for correction (A. L. Price *et al.*, 2006), and thus these covariates could immediately be added to existing models such as SKAT (Wu *et al.*, 2011) to achieve a set test that corrects for confounding. However, it is now widely accepted that there are various forms of confounders which cannot be corrected for by principal components, but for which a LMM adequately corrects (Kang *et al.*, 2010; Yu *et al.*, 2006; A. Price *et al.*, 2010), and it is for these problems that we have developed our approach.

We here focused on testing SNPs in a manner similar to SKAT (Wu *et al.*, 2011). However, it would be straightforward to also adapt FaST-LMM-Set to the approach of SKAT-O, in which the original SKAT model is in effect combined with a collapsing-type approach (Lee, Emond, *et al.*, 2012; Lee, Wu, *et al.*, 2012).

## ACKNOWLEDGEMENTS

We thank several anonymous reviewers for helpful feedback. We thank Pier Palamara for cataloguing the positions and genetic distances of SNPs in the real data, Jonathan Carlson for help with tools to manage and analyse the data, Jim Jernigan and the MSR HPC team for cluster support, and Paul de Bakker and Ciprian Crainiceanu for helpful discussions. The GAW14 data were provided by the Collaborative Study on the Genetics of Alcoholism (U10 AA008401). This study makes use of data generated by the Wellcome Trust Case-Control Consortium. A full list of the investigators who contributed to the generation of the data is available from www.wtccc.org.uk. Funding for the project was provided by the Wellcome Trust under award 076113 and 085475. We thank Hoifung Poon and Lucy Vandervende for use of their SNP-gene-disease look-up tool (http://research.microsoft.com/en-us/um/people/hoifung/SNPi/),

*Funding*: No external funding sources were used for this work.